\documentclass[8pt,aps,prl,showpacs,noshowpacs,twocolumn,latexsym,nofootinbib]{revtex4}
\usepackage[dvips]{graphicx}
\usepackage{amssymb}
\usepackage{latexsym}
\usepackage{amsmath}

\newcommand{\eq}[1]{Eq.~{\bf #1}}
\renewcommand{\Im}{\mathrm{Im}}

\begin{document}

\title{Branch-cut Singularities in Thermodynamics of Fermi Liquid Systems.}
\date{Published in \emph{Proc.~Natl.~Acad.~Sci.~USA}~\textbf{103}, 15765  (2006). }

\author{Arkady Shekhter}
\affiliation{ Department of Condensed Matter Physics, The Weizmann Institute of Science, Rehovot, 76100, Israel}
\email{E-mail:  arcadi.shehter@weizmann.ac.il.}
\author{Alexander~M.~ Finkel'stein}
\affiliation{ Department of Condensed Matter Physics, The Weizmann Institute of Science, Rehovot, 76100, Israel}

\maketitle
\newcommand{\pnassection}[1]{{\bf #1}}

{\bf The recently measured spin susceptibility of the two dimensional electron gas exhibits a strong dependence
on temperature, which is incompatible with the standard Fermi liquid phenomenology. Here we show that the
observed temperature behavior is inherent to ballistic two dimensional electrons. Besides the single-particle
and collective excitations, the thermodynamics of Fermi liquid systems includes effects of the branch-cut
singularities originating from the edges of the continuum of pairs of quasiparticles. As a result of the
rescattering induced by interactions, the branch-cut singularities generate non-analyticities in the
thermodynamic potential which reveal themselves in anomalous temperature dependences. Calculation of the spin
susceptibility in such a situation requires a non-perturbative treatment of the interactions. As in high-energy
physics, a mixture of the collective excitations and pairs of quasiparticles can be effectively described
by a pole in the complex momentum plane. This analysis provides a natural explanation for the observed
temperature dependence of the spin susceptibility, both in sign and magnitude.}

The temperature dependences of the thermodynamic quantities in the Fermi liquid have been originally attributed
to the smearing of the quasiparticle distribution near the Fermi surface \cite{Landau}. This yields a relatively
weak, quadratic in temperature, effect. A contribution of collective excitations, which in dimensions larger
than one has a small phase space has been ignored. There is a lacuna in this picture. Both the single-particle
and collective excitations are described by poles in the corresponding correlation functions. However, besides
the poles there are branch-cut singularities originating from the edges of the continuum of pairs of
quasiparticles. Such branch-cut singularities have not been given adequate attention in the theory of Fermi
liquid systems. In the Fermi liquid theory, a rescattering of pairs of quasiparticles is considered for the
description of the collective excitations which exist under certain conditions. This is not all that the
rescattering of pairs does. Regardless of the existence (or absence) of the collective modes, the excitations
near the edges of the continuum cannot be treated as independent as a consequence of the rescattering. The
thermodynamics of Fermi liquid systems is not exhausted by the contributions of the single-particle and
collective excitations. In interacting systems, as a result of the multiple rescattering, the branch-cut
singularities generate anomalous temperature dependences in the thermodynamic potential.

Motivated by recent measurements in the silicon metal-oxide-semiconductor field-effect transistors (Si-MOSFETs)
\cite{Reznikov}, we study here the temperature dependence of the spin susceptibility, $\chi(T)$, in the two
dimensional (2D) electron gas in the ballistic regime. Experiment indicates that in the metallic range of
densities and for temperatures exceeding the elastic scattering rate, $T>1/\tau_{el}$, the electrons in
Si-MOSFET behave as an isotropic Fermi liquid with moderately strong interactions.  In particular, the
Shubnikov-de Haas oscillations both without and with an in-plane magnetic field indicate clearly the existence
of a Fermi surface \cite{Pudalov,Shashkin,Kravchenko}. The only observation \cite{Reznikov} incompatible with
the simple Fermi liquid phenomenology is a surprisingly strong temperature dependence of $\chi(T)$. This
behavior occurs in a wide range of densities that rules out proximity to a  $T=0$ quantum critical point as an
explanation of the observed temperature effect. In this Report we show that such a temperature behavior of the
spin susceptibility is inherent to 2D ballistic electrons. We explain the experiment by means of anomalous
linear in $T$ terms \cite{Belitz} generated by the electron-electron (\emph{e-e}) interactions in $\chi(T)$. In
recent papers, linear in $T$ terms have been studied intensely within perturbation theory
\cite{Chubukov,Galitski,historic,Glazman,Millis}. However, these works predict the susceptibility increasing with
temperature, while the trend observed in the experiment is opposite. Taken seriously, this discrepancy indicates
that we encounter a non-perturbative phenomenon.  Here we show that a consistent treatment of the effect of
rescattering of pairs of quasiparticles in different channels provides an explanation of the observed
temperature dependence~of~$\chi$.

\pnassection{How anomalous temperature terms are generated in spin susceptibility.}
Technically, the multiple rescattering of pairs of quasiparticles is represented by ladder diagrams where each
section describes a propagation of a pair of quasiparticles between the rescattering events; see
Fig.~1. The collective excitations reveal themselves as pole singularities in the ladder
diagrams. When the pole enters into the continuum of two-particle excitations, collective excitation decay. Each
of the intermediate sections in the ladder diagrams carries two branch-point singularities which reflect the
edges of the continuum of pairs of quasiparticles. Therefore the correlation function describing a free
propagation of a pair of quasiparticles has a branch cut. The analysis of the effects of the branch-cut
singularities on temperature dependences in the thermodynamic potential is the object of this Report. In the
thermodynamic potential the contribution of the processes of multiple rescattering is given by the so-called
ring diagrams, i.e., a series of closed ladder diagrams. For the ladder diagrams, the constraints imposed by the
conservation of the momentum and energy are most effective because they are applied to a minimal number of
quasiparticles. In this way, the dominant terms are generated in the thermodynamic potential. Otherwise
summations over a large number of intermediate states smear out the singularities generated by the rescattering
processes.
\begin{figure}[h]\centerline{\includegraphics[width=0.4\textwidth]{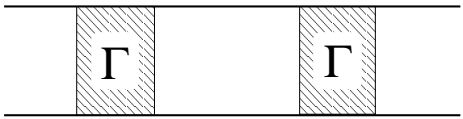}}
\caption{The diagram illustrating two rescattering events. Lines connecting the interaction amplitudes $\Gamma$ are called a ``section''. Sections represent a propagation of a pair of quasiparticles between rescatterings. Depending on the direction of the arrows (omitted here) the section may represent the propagation of particle-hole (p-h) or particle-particle (Cooper) pairs.}
\label{fig:section}
\end{figure}

We have to consider series of the ring diagrams in three different channels, i.e., in the particle-hole (p-h),
the particle-particle (Cooper), and the $2k_F$-scattering channels. The first two channels are standard for
Fermi liquid theory. The third one is mostly known in connection with the Kohn anomaly in the polarization
operator \cite{Stern}. We start by analyzing the anomalous temperature terms in the p-h channel.  Within Fermi
liquid theory, the \emph{e-e} interaction amplitude depends on the angle between the incoming and outgoing
directions of a scattered particle $\theta$ and $\theta'$ and is commonly described in terms of the angular
harmonics. To understand how the anomalous temperature terms are generated in the spin susceptibility, let us
assume for a moment that the zero harmonic, $\Gamma_0$, dominates the interaction amplitude; $\gamma=\Gamma_0$.  In
the case of a single harmonic, propagation of a p-h pair is described by the angular averaged dynamic
correlation function\footnote{We  work with the dimensionless static amplitudes known in Fermi liquid theory
\cite{Pitaevskii} as $\Gamma_n^k$, while the propagation of a p-h pair is described by the dynamic correlation
function $S(\theta)=\omega/(\omega+\Delta-qv_F\cos\theta)$; see \eq{7} in the Appendix.  Repulsion corresponds to $\Gamma_n^k>0$, and Pomeranchuk's instability is at
$\Gamma_n^k\rightarrow\infty$.}, $S_0=\langle{S}(\theta)\rangle=\omega/\sqrt{(\omega+\Delta)^2-(qv_F)^2}$, where
$\Delta=g\mu_{B}(1+\Gamma_0)H$ is the spin split energy induced by an external magnetic field $H$ and
$(1+\Gamma_0)$ describes the Fermi liquid renormalization of the $g$-factor. The function $S_0$ is imaginary
when (for a given momentum) the frequency lies within the continuum of the particle-hole excitations. The edges
of the particle-hole continuum reveal themselves in $S_0$ as a branch-point singularities. Since the position of
the branch cut $|\omega+\Delta|<v_Fq$ depends on the magnetic field, the magnetization of the electron gas
becomes sensitive to the analytical properties of the two-particle correlation function near the edge of the
continuum. The series of ladder diagrams describing the rescattering of p-h pairs generates the following term
in the magnetization (for derivation see Eqs.~(12)~and~(13) in the Appendix):
\begin{align}
&\delta{M}= \int \frac{d\omega}{2\pi} \coth\frac{\beta\omega}{2} \,\notag\\
&\Im \int_0^{\infty} \frac{qdq}{\pi}
\;\frac{\gamma\omega\tilde{\omega}}{\big[\tilde{\omega}^2-(qv_F)^2\big]\big[\gamma\omega
 +\sqrt{\tilde{\omega}^2-(qv_F)^2}\big]}\,,
\label{eq:magnetization}
\end{align}
where $\tilde{\omega}=\omega+\Delta$;  we temporarily put $g\mu_B(1+\Gamma_0)/2$ equal to one. Besides the
branch-cut singularities originating from the particle-hole continuum, the expression in \eq{1} exhibits a pole
generated by $\gamma\omega+\sqrt{\tilde{\omega}^2-(qv_F)^2}$, which determines the spectrum of the collective
excitations, i.e., the spin-wave excitations \cite{White}. Note that the expansion, either in $\gamma$ or in
$\Delta$, destroys the subtle structure of the denominator changing its analytical properties. Obviously, we
encounter a non-perturbative phenomenon.

In the case of a weak magnetic field, $\Delta<T$, the collective excitations and the continuum of particle-hole
excitations are not well-separated. Therefore, calculations of the thermodynamic quantities, e.g.,
magnetization, should be performed with care as the contributions from the collective and single-particle
excitations are not independent. Performing the $q$-integration by contours in the complex $q$-plane (one should
keep in mind that the analytical properties in the $\omega$-plane differ from that in the $q$-plane), we find
that this mixture of excitations is effectively captured by a pole in the complex momentum plane. This
finding is reminiscent of the Regge pole description of the scattering processes in high-energy physics
\cite{Gribov}. For $\Delta\neq0$, the $q$-integral is non-vanishing only when the pole in the complex $q$-plane
(a footprint of the spin-wave excitations) moves into the imaginary axis.\footnote{An alternative calculation
without referring to the complex $q$-plane is presented in Appendix.}   This move occurs within
an interval $-\Delta<\omega<-\Delta/(1+\gamma)$.  At small $\gamma$ this interval has a width $\gamma\Delta$
that, by the way, explains why in $\partial\delta{M}/\partial\Delta$ we cannot set $\Delta$ to zero. Only after
the  $q$-integration, we get for $\delta M$ an expression which (at non-zero $T$) is regular both in $\gamma$ and
$\Delta$:
\begin{equation}
\delta{M} = -\frac{\nu}{2\epsilon_F} \int_{-\Delta}^{-{\Delta}/{(1+\gamma)}}
(\omega+\Delta)\coth\frac{\beta\omega}2 d\omega\,,
\label{eq:regular}
\end{equation}
where $\nu$ is the density of states (per spin) at the Fermi surface. Expanding in $\Delta$, we obtain a linear in $T$ correction in the spin susceptibility:
\begin{equation}
\delta\chi_{\rm{p-h}} = -2\nu\frac{T}{\epsilon_F} \left(\ln\frac1{1+\gamma} + \frac{\gamma}{1+\gamma}\right)\,.
\label{eq:naive}
\end{equation}

A comment is in order here. At first glance, a linear in $T$ term in $\chi(T)$ cannot be reconciled with the
third law of thermodynamics, $S_{T\rightarrow0}=0$, in view of the Maxwell relation
$(\partial{M}/\partial{T})_H=(\partial{S}/\partial{H})_T$. This observation is the core of the statement on the textbook
level that the paramagnetic behavior with $\chi^{-1}\sim{T+\text{\emph{const}}}$  cannot exist at sufficiently low
temperature; see e.g. Ref.~\cite{Callen1960}. The well known vanishing of the coefficient of thermal expansion
at the absolute zero has the same origin. In this kind of argumentation, it is indirectly assumed that the
thermodynamic potential has a regular expansion in both of its arguments around $H,T=0$.\footnote{We
are not aware of a similar discussion of the thermal expansion coefficient (as well as elastic constants) at low
temperatures. In the context of the spin susceptibility the question has been raised  by Misawa \cite{Misawa}
who guessed (incorrectly) a non analytic form of the thermodynamic potential.} In fact, \eq{2} demonstrates
that the magnetization $\delta M=THm_{\gamma }(H/T)$ has a strong dependence on the order of limits
$H,T\rightarrow 0$. We see from \eq{2} that $\delta M\propto HT$ when $T>H,$ but for $T<H$ the temperature
dependence disappears and $\delta M\propto H^{2}$. The solution to the conflict with the third law of
thermodynamics is that the magnetic field range over which $\delta M\propto H$ shrinks to zero as $T\rightarrow
0$, and $\delta M$ acquires a non-linear in $H$ behavior outside this range. At $T\rightarrow 0$, which
unavoidably brings us into the region $T<H$, the only indisputable condition imposed by the third law is limited
to vanishing of $(\partial M/\partial T)_{H}$. Evidently, \eq{2} complies with this requirement at
$T/H\rightarrow 0$. Therefore, the existence of a linear in $T$ correction in the spin susceptibility is
legitimate and may persist down to $T\rightarrow 0$, provided that $T>H$.

\pnassection{Why spin susceptibility decreases with temperature.}
The spin susceptibility as given by \eq{3}  contradicts the experiment.  According to \eq{3} the spin
susceptibility should increase with $T$, while in the experiment it decreases.  Below we offer a resolution to
this puzzle. We would like first to indicate a subtlety in the thermodynamic potential term with two
rescattering sections (i.e., in the term proportional to $\Gamma^2$; see Fig.~2. Obviously in
the ring diagrams the number of sections is equal to the number of the interaction amplitudes). We show below
that the term $\sim\Gamma^2$ in the spin susceptibility is heavily dominated by the scattering sharply peaked
near the backward direction, $\theta-\theta'=\pi$ (throughout the paper the term "backward scattering" will be
used to refer to this process). This fact leads to far reaching physical consequences, because the diagram with
two rescattering sections dominated by backward scattering can be read in three different ways. Such a diagram
can be twisted so as to also describe the rescattering in the Cooper channel\footnote{In fact, an arbitrary
number of rescattering sections appears in the Cooper channel after such twisting, but only two of them are used
here for the extraction of the anomalous in temperature terms. The role of all other sections is to renormalize
logarithmically the {\it e-e} interaction amplitudes in the Cooper channel. In the text, we refer the term ``section''
in the Cooper channel only to those of them that generate linear in $T$ terms. This allows us to speak
simultaneously about two sections and the renormalized \emph{e-e} amplitudes without confusion.} or two sections
in the $2k_F$-scattering channel; see Fig.~2 for explanations. Therefore, overlapping of all
three channels takes place. In order to explain the sign of the effect, it is necessary to simultaneously
consider different channels and to avoid the double counting of the contributions generated by different
channels. This is the central point in our calculation of $\chi(T)$.

Before proceeding further, let us outline the consequences of the overlapping of the three channels which takes
place on the level of two rescattering sections. Instead of counting the term $\sim\Gamma^2$ in the p-h channel,
we count it within the Cooper channel where it eventually gets killed off by logarithmic renormalizations of the
interaction amplitudes. Therefore, we have to subtract the two-section term from $\delta\chi_{p-h}$ in \eq{3}
which includes it along with higher order terms. As a result of this subtraction \eq{3} has to be replaced by
Eqs.~(5) and (6) below. In the rest of this section we give the details of this procedure.
%
\begin{figure}[t]\centerline{\includegraphics[bb= 0 300 630 500,width=0.27\textwidth,angle=90]{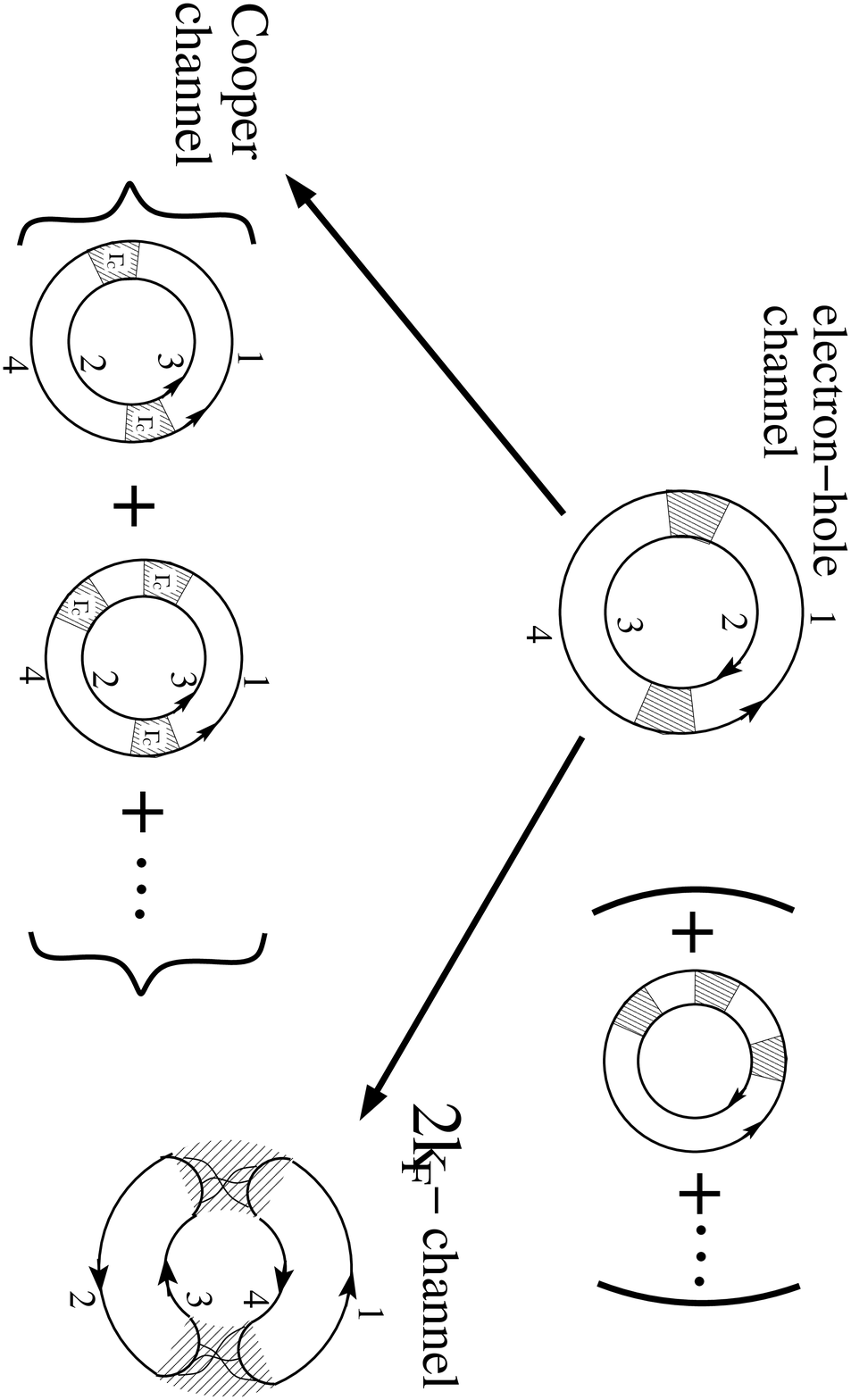}}
\caption{The diagrams at the top present the ring diagrams in the electron-hole channel. Left diagram  shows the
two-section term which is controlled by the backward scattering; the momenta in the four Green's functions are
along the same direction: $1,2\approx+\mathbf{k}_F$ and $3,4\approx-\mathbf{k}_F$. The Green functions are
numbered to keep track of them after  the rearrangement in different channels. Right diagram at the bottom shows
how the two-section term can be read in the $2k_F$-channel. Here the shaded areas represent the
interaction amplitudes in the $2k_F$-channel. The lines inside the shaded areas are drawn to clarify the spin
structure and to indicate the source of the relevant renormalizations. Left diagrams at the bottom show the
result of twisting of the two-section term into the Cooper channel. In the series of the Cooper ladder diagrams
obtained in this way only two sections (marked by numbered Green's functions) are responsible for the linear in
$T$ term in the spin susceptibility. The role of all other sections is to renormalize logarithmically the
\emph{e-e} interaction amplitudes.} \label{fig:overlap}
\end{figure}

Let us first show how the backward scattering arises in two p-h rescattering sections. This requires a
calculation of  $\chi(T)$ in which  the full harmonic content of $\Gamma(\theta-\theta')$ is included. When the
amplitudes $\Gamma_n$ and $\Gamma_m$ with different harmonics $n$ and $m$ are involved, the propagation of the
p-h pair between the rescattering events is described by the dynamic correlation functions
$S_{nm}=[(\tilde{\omega}-\sqrt{\tilde{\omega}^2-(v_Fq)^2}\,)/v_Fq]^{|n-m|}S_0$. Despite the nontrivial
dependence of $S_{nm}$ on the harmonics indices, the contribution to  $\chi(T)$ in the second order of
\emph{e-e} interaction amplitudes acquires  a very simple form:
\begin{equation}
\delta\chi_{(2)} = \nu\frac{T}{\epsilon_F} \sum_{nm} (-1)^{n+m}\Gamma_n\Gamma_m\,.
\end{equation}
Because $\sum_{n} (-1)^{n}\Gamma_n$ is equal to the backward scattering amplitude $\Gamma(\pi)$, this contribution reduces to  $\delta\chi_{(2)} = ({T}/{\epsilon_F})\nu \Gamma^2(\pi)$.\footnote{A calculation with the use of angular harmonics has been performed in \cite{Millis} for the anomalous terms in the specific heat; it also leads to the backward scattering amplitude~$\Gamma(\pi)$.} We have checked that exactly the same result can be obtained by the calculation of two rescattering sections in the Cooper channel, or in the $2k_F$-channel. In the calculation of the Cooper channel, we use the angular harmonics of the particle-particle correlation functions. Once again, despite the nontrivial dependence of these correlation functions on their harmonics indices, the result reduces to the backward scattering amplitude; details will be published elsewhere \cite{elsewhere}. Moreover, this calculation yields the same coefficient as in \eq{4}. In the case of the $2k_F$-channel the presence of $\Gamma(\pi)$ in \eq{4} is evident, but one has to check the coefficient. On the level of two rescattering sections the contributions generated in three channels coincide (i.e., in $\delta\chi_{(2)}$ all three channels overlap),  as we described above.

We now analyze the problem of the renormalizations of the linear in $T$ terms. It is easy to check that unlike the case of one-dimensional electrons \cite{BGD1966},  the higher order terms in the $2k_F$-scattering channel are not important in 2D. Therefore, we will not discuss this channel further and concentrate on the interplay between the other two channels. Up to this point, the interaction amplitudes have played a rather passive role in our calculations. The peak near the backward scattering direction has been generated by the dynamic correlation functions describing the propagation of pairs of particles in each of the channels. The interaction amplitudes have simply supplied a featureless coefficient $\Gamma(\pi)$ in the two-section term. To understand the true role of the \emph{e-e} interaction in the anomalous temperature corrections we have to abandon the central assumption of the microscopic Fermi liquid theory that different sections in the ladder diagrams are independent. Indeed, when the rescattering is dominated by the backward scattering, a strong dependence of the interaction amplitude on its arguments in the p-h channel emerges from the logarithms in the Cooper channel (this is a weak version of the parquet known for one-dimensional electrons \cite{BGD1966}). In view of this circumstance, in the case of two rescattering sections we have to take into consideration the dependence of the scattering amplitude $\Gamma(\mathbf{p},\mathbf{-p+q+k},\mathbf{-p+k},\mathbf{p+q})$ on the arguments $\mathbf{q}$ and $\mathbf{k}$. We resolve the problem of the logarithms by moving the term with two rescattering sections to the Cooper channel where the logarithmic renormalizations originate. This move is possible because the terms with two sections in different channels coincide. Note also that as a result of moving the two-section term to the Cooper channel we avoid the double counting of the two-section term in three different channels. After this step, as we discussed earlier, the contribution to the spin susceptibility from the p-h channel becomes
\begin{equation}
\delta\chi' = \delta\chi_{\rm{p-h}} - \delta\chi_{(2)}\,.
\end{equation}

We now  consider the contribution to the spin susceptibility from the Cooper channel. The rescattering in the Cooper channel leads to the logarithmic renormalizations of the interaction amplitudes $\Gamma_n^C(T)=\Gamma_n^C/(1+\Gamma_n^C\ln\epsilon_F/T)$  where $\Gamma_n^C$ are harmonics of the amplitude in the Cooper channel. At sufficiently small temperatures,  the repulsive amplitudes, $\Gamma_n^C>0$, vanish as $\Gamma_n^C(T)\approx1/\ln(\epsilon_F/T)$. (We do not consider here the developing of the instability for the attractive amplitudes \cite{elsewhere} as it is most likely blocked by the disorder in the system studied in Ref.~\cite{Reznikov}.) Therefore, the linear in $T$ terms generated in the Cooper channel are suppressed at low temperatures. Coming back to the discussion preceding \eq{5}, we now see that the logarithmic renormalization of the amplitudes in the term with two rescattering sections in the p-h channel results in full elimination of this term at low enough temperatures. Therefore, for the repulsive \emph{e-e} interaction when only zero harmonic is kept, the temperature dependence of the spin susceptibility is given by
\begin{equation}
\delta\chi = -2\nu\frac{T}{\epsilon_F}\left(\gamma^2/2 + \ln\frac1{1+\gamma} + \frac{\gamma}{1+\gamma}\right)\,.
\end{equation}
In contrast to the previous calculations, this result provides
the sign of the temperature dependence of the spin susceptibility which coincides with that observed
experimentally \cite{Reznikov}. The expression in \eq{6} has been obtained by summation of the ladder diagrams
in two channels and taking into consideration the overlap of the two-section term. In this way we resolve the
puzzle of the sign of the temperature trend in $\chi(T)$.

We next note that the intervention of the Cooper renormalizations in the p-h channel is effective only for the
term with two rescattering sections. We have checked that the situation with a dominant role of the backward
scattering is not general and it does not occur for terms with more than two rescattering sections. A direct
calculation of the term with three interaction amplitudes
$\delta\chi_{(3)}=(T/\epsilon_F)\nu\int\alpha(\theta_1\theta_2\theta_3)\Gamma(\theta_1-\theta_2)\Gamma(\theta_2-\theta_3)\Gamma(\theta_3-\theta_1)d\theta_1d\theta_2d\theta_3$,
performed with the use of the methods sketched above, shows that there is only a weak (logarithmic) singularity
near the backward scattering. This is far weaker than the sharp $\delta$-function peak near the backward
direction, $\theta-\theta'=\pi$, in the case of two rescattering sections. It is therefore safe to conclude
that, unlike the case of the two-section term, the logarithmic renormalizations are ineffective for three and
more sections in the p-h channel.
\begin{figure}[h]\centerline{\includegraphics[width=0.5\textwidth]{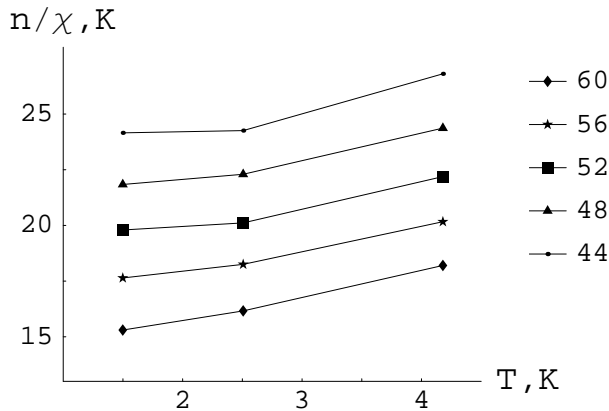}}
\caption{The experimental data for $n/\chi$ expressed in $K^{\circ}$ and depicted as a function of temperature.  $\chi$ is the spin susceptibility determined from $H=0.7 T$ spin magnetization; $n$ is the density of the 2D electron gas in Si-MOSFET. Numbers in the inset are density of the corresponding curves in units of $10^{10}cm^{-2}$. Only a fragment of the data of Ref.~\cite{Reznikov} corresponding to the ballistic electrons is presented. [Data courtesy of M. Reznikov (Technion, Haifa), used with permission.]}
\label{fig:reznikov} 
\end{figure}

\pnassection{Relation to experiment.}
Finally, let us discuss the result of our analysis in connection with the measurement of the spin
susceptibility in Si-MOSFET \cite{Reznikov}. In Fig.~3 the data for a quantity  $n/\chi$  are
presented, where $n$ is the density of the 2D electron gas. We focus here on the curves corresponding to the
ballistic range of the densities.  These curves exhibit a noticeable rise with temperature at $T\geq1.5K$, which
is too strong for the conventional Fermi liquid theory; the conventional Fermi liquid temperature dependence
will be invisible on scales used in the plot of Fig.~3. The data correspond to the range of
densities and temperatures where the transport is ballistic. The rising curves in this plot indicate that the
spin susceptibility decreases with temperature.  We assume here that this temperature dependence is due to the
term $\delta\chi$  given by \eq{6}, which we multiply by the factor $4$ to account for two valleys. At lower
temperatures the discussed effect of the anomalous temperature corrections is cut-off by disorder.  One can
expand $n/\chi$  with respect to the temperature corrections: $\delta(n/\chi)=T\,f(\gamma)$ , where $f(\gamma)=
\gamma^2/2 - \ln(1+\gamma) + \gamma/(1+\gamma)$. The modification of the spin susceptibility by the Stoner
factor $(1+\Gamma_0)$ drops out from $\delta(n/\chi)$. It is cancelled by the two factors $(1+\Gamma_0)$ in
$\delta\chi$ ignored so far because in the definition of $\Delta$ the combination $g\mu_B(1+\Gamma_0)/2$ has
been put equal to one. The main advantage of the combination  $\delta(n/\chi)$ is that its temperature
dependence is determined by the dimensionless interaction amplitudes only.  In the discussed range of densities,
parameter $r_s$ is about $3\div4$ ($r_s$ is the ratio of the energy of the \emph{e-e} interaction to the kinetic
energy).  Therefore, one may expect the dimensionless interaction amplitudes to have a magnitude $\sim1$. Perhaps even a few leading harmonics may be involved.  For $n\neq0$ harmonics enter in pairs, $\Gamma_n=\Gamma_{-n}$, and consequently $f(\gamma)$ should be slightly modified because of mixing between $\Gamma_n$ and $\Gamma_{-n}$; see Appendix for details. When the amplitude $\gamma\sim1$  the function $f(\gamma)$ is of order unity (e.g., for $\gamma=1.5$ it is equal to $0.7$). The slope of the curves
$\delta(n/\chi)$ presented in Fig.~3 is also $\sim1$, i.e., of the same order of magnitude.
Together these facts support our conclusion that   at low temperature  the sign and the magnitude of the temperature
dependence of the spin susceptibility  can be explained by the theory of the anomalous corrections presented in
this Report. At temperatures comparable with the Fermi energy the logarithmic suppression of the
interaction amplitudes in the $\delta\chi_{(2)}$-term should become ineffective. If so, when
$1/\ln(\epsilon_F/T)\gtrsim\gamma$ the temperature dependence will change sign leading to a non-monotonic
spin susceptibility. Unfortunately, the temperature range of the existing measurement,
$T/\epsilon_F\lesssim0.1$, does not allow to verify this consequence of our theory.

To conclude, the thermodynamics of Fermi liquid systems is not exhausted by the contributions of the
single-particle and collective excitations.  These two types of the excitations are described by the poles in
the corresponding correlation functions. However, the theory of Fermi liquid systems is not complete without
consideration of the branch-cut singularities. In interacting systems, as a result of the rescattering of
quasiparticles, the branch-cut singularities generate non-analyticities in the thermodynamic potential which
reveal themselves in anomalous temperature dependences. The observed temperature dependence in the spin
susceptibility of the 2D electron gas can be explained in this way. The mechanism determining the sign of the anomalous terms in the spin susceptibility discussed here may have implications for the physics near the quantum critical point at the ferromagnetic instability.

\appendix
\section*{Appendix}

Here we present the details of the calculation of the anomalous temperature
corrections to the spin susceptibility originating from the particle--hole (p--h)
channel. The propagation of the pair of the quasiparticles with the opposite
spin projections in the p--h channel is described (see \S ~17 in ref.~{13}  in
the main text) by the two-particle correlation function $[GG]_{q,\omega
,\Delta }=\nu \,{\delta\epsilon}/(\omega-\delta\epsilon)$, where $\delta \epsilon =\epsilon _{\mathbf{p+q}\uparrow }-\epsilon _{\mathbf{p}%
\downarrow }=\mathbf{v}_{F}\mathbf{q}-\Delta $. Here $\Delta =2h$ is the
relative shift of the chemical potential equal to the Zeeman energy
splitting; $h=(g\mu _{B}/2)B$. It is convenient to single out the dynamic
part of this correlation function:
\begin{align}
&[GG]_{q,\omega ,\Delta }=-\nu +\nu S(\theta )_{q,\omega ,\Delta
};\quad \quad \notag\\
&S(\theta )_{q,\omega ,\Delta }=\frac{\omega }{\omega +\Delta
-v_{F}q\cos \theta }\,.  \label{eq:zerosound}
\end{align}%
In this work, the static part of $[GG]_{q,\omega ,\Delta }$ is absorbed in
the static Fermi-liquid amplitudes $\Gamma ^{k}$ and will be not considered
further. The propagation of a p--h pair is described by the dynamic
correlation function $S(\theta )$. (From now on, we will omit $v_{F}$ in the
product $v_{F}q$ and consider $q$ as measured in energy units.)

Let us first calculate the contribution from the two rescattering sections
in the p--h channel to the anomalous term in the thermodynamic potential:
\begin{eqnarray}
\Omega _{2}(\Delta ) &=&-(\nu /\epsilon _{F})\sum_{n,m}\frac{\Gamma
_{n}^{k}\Gamma _{m}^{k}}{4}\int \frac{qdq}{2\pi }\int {d\omega }\coth \frac{%
\beta \omega }{2}  \notag \\
&\times &\,\Im \big\{S_{n-m}(q,\omega ,\Delta )S_{m-n}(q,\omega ,\Delta )%
\big\}\,.  \label{eq:zssecond}
\end{eqnarray}%
Because the amplitudes with different harmonics $\Gamma _{n}$ and $\Gamma _{m}$
are involved, the propagation of the p--h pair is described by the angular
harmonics of the two-particle correlation function, $S_{n-m}=\int ({d\theta }%
/{2\pi })S(\theta )e^{i(n-m)\theta }$, where
\begin{equation}
\,S_{0}=\frac{\omega }{\sqrt{{\tilde{\omega}}^{2}-q^{2}}}\qquad S_{l\neq
0}=\left( \frac{{\tilde{\omega}}-\sqrt{{\tilde{\omega}}^{2}-q^{2}}}{q}%
\right) ^{|l|}S_{0}\,.  \label{eq:azsharmonics}
\end{equation}%
Here we introduce ${\tilde{\omega}}={\omega }+\Delta $. Integration over $q$
can be done by contours in the complex $q$ plane. This leads to the
following frequency integral:
\begin{equation}
\,\Omega _{2}(\Delta )=(\nu /\epsilon _{F})\sum_{n,m}\frac{(-1)^{n-m}\Gamma
_{n}^{k}\Gamma _{m}^{k}}{16}\int d{\omega }\,{\omega }^{2}\coth \frac{\beta
\omega }{2}\mathrm{sign}{{\tilde{\omega}}}\,.  \label{eq:zsseconda}
\end{equation}%
Performing the frequency integration for the spin susceptibility $\chi
=-\partial ^{2}\Omega /\partial {h}^{2}$ we come to Eq.~{\bf{4}}  in the main
text:
\begin{equation}
\,\delta \chi _{(2)}=\nu \frac{T}{\epsilon _{F}}\sum_{n,m}{(-1)^{n-m}\Gamma
_{n}^{k}\Gamma _{m}^{k}}\,.  \label{eq:zsresult}
\end{equation}

We now calculate the ladder diagrams in the p--h channel. We start with the zero harmonic; $\Gamma_0=\gamma$. For completeness we
will do it in two different ways. The contribution to the thermodynamic
potential is equal to
\begin{align}
&\delta\Omega (\Delta )\label{eq:ringa}\\
&=-(\nu /2\epsilon _{F})\int d{\omega }\coth \frac{%
\beta {\omega }}{2}\,\,\frac{qdq}{2\pi }\,\Im \,\ln \frac{1}{1+\gamma {S}%
_{0}(q,\omega ,\Delta )}  \notag \\
&=-(\nu /2\epsilon _{F})\int {d{\omega }}\coth \frac{\beta {\omega }}{%
2}\,\,\frac{qdq}{2\pi }\,\Im \,\ln \frac{\sqrt{{\tilde{\omega}}^{2}-q^{2}}}{%
\gamma \omega +\sqrt{{\tilde{\omega}}^{2}-q^{2}}}\,.  \notag
\end{align}%
Here we assume that the frequency $\omega $ is slightly shifted above the
real axes. Eq.~{\bf{1}} in the main text for the magnetization $\delta
M=-\partial \delta \Omega /\partial {h}$ follows immediately from this
expression. To proceed further, we observe that ${\tilde{\omega}}$ and $q$
enter in \eq{\ref{eq:ringa}}  only through the combination ${\tilde{\omega}}^{2}-q^{2}$. Therefore the expression inside the integral vanishes under the
action of ${\tilde{\omega}}^{-1}\partial /\partial {{\tilde{\omega}}}+q^{-1}{\partial/\partial{q}}$. We use this to write the magnetization $M=-\partial\Omega/\partial {h}$ as an integral of the full derivative
\begin{align}
&\delta {M}=(\nu /\epsilon _{F})\int \frac{d{\omega }}{2\pi }\coth \frac{\beta {\omega }}{2}\,{\tilde{\omega}}\,\notag\\
&\int_{0}^{\infty }{dq}{\frac{\partial}{\partial {q}}}\left(-\Im\ln\frac{\sqrt{{\tilde{\omega}}^{2}-q^{2}}}{\gamma\omega+\sqrt{{\tilde{\omega}}^{2}-q^{2}}}\right) \, .  \label{eq:ringF}
\end{align}
Collecting the contribution at the lower limit of the $q$ integral, we
obtain
\begin{equation}
\delta M=(\nu /\epsilon _{F})\int \frac{d{\omega }}{2\pi }\coth \frac{\beta {%
\omega }}{2}\,{\tilde{\omega}}\,\Im \,\ln \frac{{\tilde{\omega}}}{\gamma
\omega +{\tilde{\omega}}}\,.
\end{equation}
Here we have used that $\sqrt{{\tilde{\omega}}^{2}-q^{2}}|_{q\rightarrow 0}={%
\tilde{\omega}}$, which corresponds to the correct analytical structure of
the square root function in $S_{0}$ and $S_{l}$. For $\omega $ slightly
above the real axis we have
\begin{align}
&\mathrm{for}\,\,\gamma >0\quad \Im \ln \frac{{\omega }+\Delta }{(\gamma
+1)\omega +\Delta }=\left\{
\begin{array}{cc}
-\pi , & -\Delta <{\omega }<-\frac{\Delta }{1+\gamma } \\
0, & \mathrm{otherwise}%
\end{array}%
\right\} \,.  \notag \\
&\mathrm{for}\,\,\gamma <0\quad \Im \ln \frac{{\omega }+\Delta }{(\gamma
+1)\omega +\Delta }=\left\{
\begin{array}{cc}
\pi , & -\frac{\Delta }{1+\gamma }<{\omega }<-\Delta  \\
0, & \mathrm{otherwise}%
\end{array}%
\right\} \,.  \label{eq:logrationeg}
\end{align}%
As a consequence of taking an imaginary part, the frequency integration is
restricted to a narrow frequency interval around zero. As a result we obtain
Eq.~{\bf{2}}  of the main text:
\begin{equation}
\delta {M}=-\frac{\nu }{2\epsilon _{F}}\int_{-\Delta }^{-{\Delta }/{1+\gamma
}}(\omega +\Delta )\coth \frac{\beta \omega }{2}d\omega \,.
\label{eq:regular}
\end{equation}

It is instructive to reproduce the same result by a more powerful (but also
more delicate) method of the integration by contours in the complex $q$ plane. We return to the expression for magnetization $\delta{M}$:
\begin{align}
&\delta{M} = (\nu/\epsilon_F) \int {d{\omega}}\coth\frac{\beta{\omega}}{2}
\notag\\
&\Im \, \int_0^{\infty} \frac{qdq}{2\pi} \frac{\gamma{\omega}{\tilde{\omega}%
}}{({\tilde{\omega}}^2-q^2)(\gamma\omega+\sqrt{{\tilde{\omega}}^2-q^2})}.
\label{eq:complexringF}
\end{align}
An important property of this expression is that apart from the branch cut
on the real axes it also has  poles in the complex $q$ plane originating from
zeros of $\gamma\omega+\sqrt{{\tilde{\omega}}^2-q^2}$. Solving the equation $\gamma\omega+\sqrt{{\tilde{\omega}}^2-q^2}=0$ we find that the poles never
appear inside the branch cut in the complex $q$ plane (the branch-cuts are along the real axis, covering $|q|>\tilde{\omega}$). The poles are either
somewhere on the real axes between $-\tilde{\omega}$ and $\tilde{\omega}$ or
appear as a pair on the imaginary axis. For $\gamma>0$ the poles exist if $%
-\Delta<\omega<0$. They are imaginary for $-\Delta<\omega<-\Delta/(1+\gamma)$%
. [For $\gamma<0$ the imaginary poles exist for $-\Delta/(1+\gamma)<\omega<-%
\Delta$.] We see that the conditions that the poles are on the imaginary
axes lead to the same intervals as in \eq{\ref{eq:logrationeg}}  above.

The expression under the integral in \eq{\ref{eq:complexringF}}  is an odd function of $q$.
This allows us to rewrite the $q$ integral as a contour integral in the
complex $q$ plane. The contour consists of two lines going in the opposite
direction above and below the real axis. More specifically, the part of the
contour below the real axis goes in the positive direction when $\tilde{%
\omega}>0$ and in the negative direction when $\tilde{\omega}<0$. Thus,
\begin{align}
&\delta{M} = (\nu/4\epsilon_F) \int {d{\omega}} \coth\frac{\beta{\omega}}{2} \,
\notag\\
&\int_C \frac{qdq}{2\pi i} \frac{\gamma{\omega}{\tilde{\omega}}}{({%
\tilde{\omega}}^2-q^2)(\gamma\omega+\sqrt{{\tilde{\omega}}^2-q^2})}\,.
\label{eq:complexringFF}
\end{align}
When there are no poles, or they are present but located on the real axis $q$,
the integral vanishes. (The contour $C$ can be deformed to a big circle
where the function under the integral drops as $\sim\!\!1/q^2$.) However, the
integral does not vanish if the poles are on the imaginary axis in the
complex $q$ plane. This occurs only for the frequency intervals discussed
above. By deforming the contour and taking the residue we reproduce the
Eq.~{\bf{2}}  in the main text.

So far, only one harmonic in the scattering amplitude has
been iterated within the ring diagrams. Moreover, for the purpose of clarity
it has been assumed that zero harmonic amplitude $\Gamma _{0}$ is dominant
leading to Eq.~{\bf{6}}  in the main text. When $n\neq 0,$
harmonics enter in pairs because harmonic amplitudes for $\pm n$ are equal, $%
\Gamma _{n}=\Gamma _{-n}$. When different harmonics are involved, the
segments representing the iterated harmonics $\hat{\gamma}_{\pm n}=\Gamma
_{n}/(1+\Gamma _{n}\omega /\sqrt{\tilde{\omega}^{2}-q^{2}})$ have to be
connected by a correlation function $S_{n-m}$ which represents the
''transition section'' from $\Gamma _{n}$ to $\Gamma _{m}$. For a particular
case of a pair of harmonics $\Gamma _{\pm n}$, the transition section is $%
S_{2n}$. The summation of the ring diagrams for $n\neq 0$ can be performed
by counting how many times the section $S_{2n}$ appears
\begin{eqnarray}
&\,&\delta {\Omega }(\gamma _{n},\gamma _{-n};\Delta )=-(\nu /2\epsilon
_{F})\int {d\omega }\coth \frac{\beta \omega }{2}
\notag \\
&\,&\quad \times \Im \int \frac{qdq}{2\pi }\,\ln \frac{\hat{\gamma}_{n}\hat{\gamma}_{-n}}{1-\hat{\gamma%
}_{n}S_{2n}\hat{\gamma}_{-n}S_{2n}}.  \label{eq:all-two-harmonics}
\end{eqnarray}

In the following we will limit ourselves to the first mixing term in $\delta
{\Omega }$ in which the transition section $S_{2n}$ appears two times; this
is reasonable when $\Gamma _{n}\lesssim 1$. For clarity, we will denote $%
\Gamma _{\pm n}$ as $\gamma _{\pm }$. Then,
\begin{align}
&\delta {\Omega }(\Delta )=-(\nu /2\epsilon _{F})\int {d\omega }\coth
\frac{\beta \omega }{2}\,\Im \int \frac{qdq}{2\pi }  \notag \\
&\times \,\frac{\gamma _{+}\omega }{\gamma _{+}\omega +\sqrt{\tilde{%
\omega}^{2}-q^{2}}}\frac{\gamma _{-}\omega }{\gamma _{-}\omega +\sqrt{\tilde{%
\omega}^{2}-q^{2}}}\left( \frac{\tilde{\omega}-\sqrt{\tilde{\omega}^{2}-q^{2}%
}}{\tilde{\omega}+\sqrt{\tilde{\omega}^{2}-q^{2}}}\right) ^{2n}.
\label{eq:mixing-two-harmonics}
\end{align}%
The $q$ integral is evaluated by contours in the complex $q$ plane
\begin{align}
&\delta {\Omega }(\Delta )=\nu \Delta ^{2}\frac{T}{2\epsilon _{F}}\gamma
_{+}\gamma _{-}+\delta {\Omega }_{\pm }(\Delta ),  \label{eq:mix} \\
&\delta {\Omega }_{\pm }(\Delta )=\frac{1}{4}(\nu /\epsilon _{F})\gamma
_{+}\gamma _{-}  \\
&\times\Big[ \int_{-\Delta }^{-\Delta /1+\gamma _{+}}{d\omega }\,\omega
^{2}\coth \frac{\beta \omega }{2}\frac{\gamma _{+}}{(\gamma _{-}-\gamma _{+})%
}\left( \frac{\tilde{\omega}+\gamma _{+}\omega }{\tilde{\omega}-\gamma
_{+}\omega }\right) ^{2n}\,\notag\\
&\qquad\qquad\qquad+(\gamma _{+}\leftrightarrow \gamma _{-})\Big] .\notag
\end{align}%
The first term $\delta {\Omega }(\Delta )$ is a part of the two section term
which reduces to the square of the backward scattering amplitude. It gets
killed by the logarithmic renormalizations in the Cooper channel as it has
been discussed in the main text; we will not keep this term anymore. The
other term in \eq{\ref{eq:mix}}  is determined by the poles of $1/(\gamma
_{\pm }\omega +\sqrt{\tilde{\omega}^{2}-q^{2}})$ when they are on the
imaginary axis. After passing to a new variable $x=(\omega /\Delta
+1)(1+\gamma _{+})/\gamma _{+}$ we obtain in the limit $\beta \Delta \ll 1$:
\begin{equation}
\delta {\Omega }_{\pm }(\Delta )=\nu \Delta ^{2}\frac{T}{2\epsilon _{F}}%
\gamma _{+}\gamma _{-}\frac{\phi _{n}(\gamma _{+})-\phi _{n}(\gamma _{-})}{(\gamma _{+}-\gamma _{-})},
\end{equation}%
where the function describing the mixing of two harmonics is
\begin{equation}
\phi _{n}(\gamma )=\frac{\gamma^{2}}{1+\gamma }\,\int_{0}^{1}{dx}\,\left( 1-\frac{%
\gamma x}{1+\gamma }\right) \,\left( \frac{x-1}{1+\frac{1-\gamma }{1+\gamma }\, x}\right) ^{2n}.
\end{equation}%
The last factor reduces the numerical value of $\phi _{n}$ so that the correction from the mixing of harmonics
is noticeable mostly for first few harmonics with $n\neq1$. Using the fact that the magnitude of
amplitudes $\gamma _{+}$ and $\gamma _{-}$ is equal, we obtain
\begin{equation}
\delta {\chi }_{\pm }=-2\nu (T/\epsilon _{F})\;\Gamma _{n}^{2}\frac{d\phi
_{n}(\Gamma _{n})}{d\Gamma _{n}}.
\end{equation}
Finally, we obtain for a pair of non-zero harmonics $\pm n$
\begin{equation}
\delta {\chi }_{n\neq 0}=-2\nu (T/\epsilon _{F})\Big[f(\Gamma _{n})+f(\Gamma
_{-n})+\Gamma _{n}^{2}\frac{d\phi
_{n}(\Gamma _{n})}{d\Gamma _{n}}\Big].
\end{equation}%
All terms on the right-hand side start with $\Gamma_n^3$. The last term describes the modification of the temperature term in the spin
susceptibility for non-zero harmonics because of the mixing of $\Gamma _{n}$
and $\Gamma _{-n}$; compare with \eq{6}  in the main text. At $\Gamma=1$, the value $2f(\Gamma=1)\approx0.6$, whereas for $n=1; 2; 10$ the corresponding values of $\Gamma ^{2}({d\phi_{n}(\Gamma)}/{d\Gamma})$ are $0.24; 0.16; 0.04$.

A similar analysis can be performed for the terms with three (or more)
interaction amplitudes. We check in this way that for the higher order terms
in the p--h channel the scattering does not reduce to the backward
scattering. Therefore, there is no overlap with the Cooper channel, and the
intervention of the Cooper channel is irrelevant here.

{\bf Acknowledgements:} We thank M. Reznikov and C. Varma for valuable discussions. A.F. is supported by the Minerva Foundation.

\end{document}